\documentclass[aps,prl,showpacs,twocolumn,letterpaper,nobibnotes]{revtex4-1}
\usepackage{amssymb}
\usepackage{amsmath}
\usepackage{graphicx}
\usepackage{url}
\newcommand\ket[1]{\ensuremath{|#1\rangle}}
\newcommand\bra[1]{\ensuremath{\langle#1|}}
\newcommand\mean[1]{\ensuremath{\left<#1\right>}}
\begin{document}
\title{Vacuum spin squeezing}
\author{Jiazhong Hu}
\author{Wenlan Chen}
\author{Zachary Vendeiro}
\author{Alban Urvoy}
\author{Boris Braverman}
\author{Vladan Vuleti\'{c}}
\affiliation{Department of Physics and Research Laboratory of Electronics, Massachusetts Institute of Technology,
Cambridge, Massachusetts 02139, USA}
\begin{abstract}
We investigate the generation of entanglement (spin squeezing) in an optical-transition atomic clock through the coupling to a vacuum electromagnetic field that is enhanced by an optical cavity. We show that if each atom is prepared in a superposition of the ground state and a long-lived electronic excited state, and viewed as a spin-1/2 system, then the collective vacuum light shift entangles the atoms, resulting in a squeezed distribution of the ensemble collective spin. This scheme reveals that even a vacuum field can be a useful resource for entanglement and quantum manipulation. The method is simple and robust since it requires neither the application of light nor precise frequency control of the ultra-high-finesse cavity. Furthermore, the scheme can be used to implement two-axis twisting by rotating the spin direction while coupling to the vacuum, resulting in stronger squeezing.
\end{abstract}
\maketitle

Accurate time and frequency measurements are important for a variety of applications. High precision clocks \cite{clock3,clock6,clock8,clock9,clock10,clock1,clock2,clock7} enable applications such as position locating, high-resolution measurement of atomic and molecular transitions \cite{transition1,transition2,Chen08}, and precision sensing of gravity \cite {gravity1,gravity2,GravityG}. In addition, it has been proposed that atom interferometry with clock-transition atoms can be used for long-baseline gravitational wave detection \cite{atomicgravity}.

Atomic clocks represent one of the most impressive advances in technology of the last decades \cite{OA}. By taking advantage of ultranarrow optical transitions \cite{clock1,clock2,clock3,clock6,clock7,clock8,clock9,clock10} in ensembles of many trapped atoms, the accuracy of time measurements has been improving continuously, and has now reached a fractional stability in the $10^{-18}$ range \cite{clock7,clock8,clock10}. Such clocks now operate near the standard quantum limit (SQL) that is associated with the quantum projection noise for measurements on independent particles \cite{UC}. 

The SQL can be overcome by incorporating an entangled state of many atoms as an input state to the standard Ramsey sequence \cite{PhysRevA.46.R6797,squeeze,sclock,20dB,My,sclock,James,Polzik1,Mitchell1,nonG,Chapman1,Bohnet1297,Klempt1,Treutlein1}. A particularly simple and robust many-atom entangled state is a squeezed spin state \cite{PhysRevA.46.R6797} where the noise in the phase quadrature is reduced at the expense of increased noise in the population quadrature. Two experimentally demonstrated methods used for generating spin squeezing in large atomic ensembles are atomic collisions \cite{nonG,Chapman1,Bohnet1297,Klempt1,Treutlein1} and atom-light interaction \cite{James,My,20dB,Polzik1,Mitchell1}. In 2010, an atomic clock operated by 3~dB below the SQL has been demonstrated \cite{sclock}, and recently a state squeezed by up to 18~dB, corresponding to a reduction of variance by a factor of 60, has been observed \cite{20dB}. 
\begin{figure}[htbp]
\begin{center}
\includegraphics[width=3.7in]{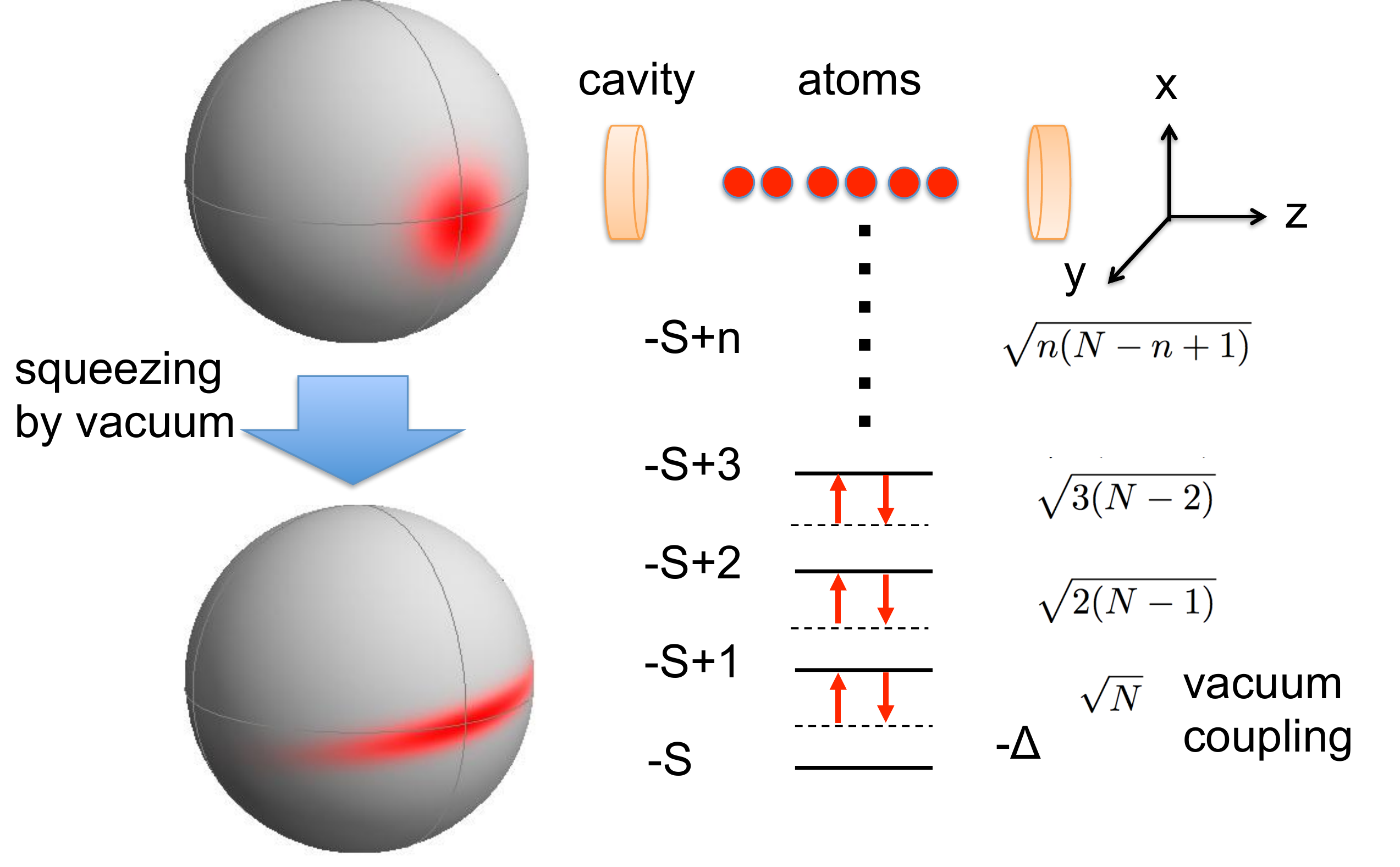}
\caption{Principle of vacuum spin squeezing on an optical clock transition. $N$ two-level atoms are trapped in a high finesse optical cavity. The cavity enhanced vacuum is coupling the atomic transition, $\ket{\uparrow}$ $\rightarrow$ $\ket{\downarrow}$, with a detuning $\Delta$. After the atomic ensemble is initialized into a coherent spin state near the equator, the atomic spin distribution begins to spontaneously squeeze. The cavity coupling is detuned from the atomic resonance by $\Delta$, so that virtual photon emission into the cavity and photon reabsorption leads to a light shift of the atomic levels. Since the coupling to the cavity for different Dicke states varies with $S_z$, each Dicke state experiences a different nonlinear light shift. This results in an $S_z$-dependent state rotation that is equivalent to one-axis twisting \cite{squeeze}.}
\label{Fig1}
\end{center}
\end{figure}

While most optical methods generates conditional spin squeezing by measurement, cavity feedback squeezing is an experimentally demonstrated \cite{sclock} deterministic and unconditional spin squeezing method that uses light to generate an effective atom-atom interaction. In this scheme, for incident light tuned to the slope of the cavity resonance, the tuning of the cavity by the atomic spin and the associated change in intracavity intensity establish quantum correlations between different components of collective atomic spin. Cavity feedback squeezing, as well as measurement squeezing methods \cite{James,20dB}, require extremely good cavity length and laser frequency stabilization, so that the intracavity intensity is determined by the quantum noise of the atoms and the light, rather than technical cavity or laser noise. All spin squeezing methods so far \cite{Ma201189} have generated spin squeezing on microwave or radiofrequency transitions in the atoms' electronic ground states, leaving the application to optical transitions to be explored.

In this Letter, we show that an electromagnetic vacuum can also induce an effective spin dependent interaction between distant atoms. Considering clock atoms with a narrow optical transition such as in strontium \cite{clock3,clock6,clock9}, calcium \cite{clock8}, aluminum \cite{clock1}, or ytterbium \cite{clock2,clock7}, we show that the collective light shift arising from the cavity vacuum field entangles all atoms in the ensemble. This vacuum spin squeezing thus reveals a surprising non-trivial property of vacuum, namely, that virtual photon emission into a vacuum mode by a collection of particles can result in entanglement between the particles. The method is robust in that the cavity resonance frequency need not be maintained very precisely compared to its linewidth. By adding a rotation of the atomic spin during the vacuum spin squeezing, the effective one-axis twisting Hamiltonian \cite{squeeze} can be converted into two-axis twisting, resulting in stronger and faster squeezing.

We consider $N$ two-level atoms trapped inside an optical cavity using a magic-wavelength trap \cite{Magic} (Fig.~\ref{Fig1}). The excited state $\ket{e}=\ket{\uparrow}$ decays to ground state $\ket{g}=\ket{\downarrow}$ by emitting light at frequency $\omega_0$. The cavity field is coupling the two states $\ket{\downarrow}$ and $\ket{\uparrow}$ with a detuning $\Delta$ ($|\Delta|\ll\omega_0$). The Hamiltonian of the composite atom-cavity system (ignoring free space emission and cavity decay for the moment) is described by a standard Tavis-Cummings model \cite{TC1,TC2,TC3}, which is written as
\begin{equation}
H/\hbar=\sum^N_{j=1} \omega_0 \sigma^z_j/2+(\omega_0+\Delta)c^\dagger c+\sum_{j=1}^N g(c^\dagger \sigma^-_j+c \sigma^+_j),
\end{equation}
where $c^\dagger$ ($c$) is the creation (annihilation) operator of the cavity field, $2g$ is the single photon Rabi frequency, and $\sigma^z_j=\ket{\uparrow}_j\bra{\uparrow}_j-\ket{\downarrow}_j\bra{\downarrow}_j$, $\sigma^+_j=\ket{\uparrow}_j\bra{\downarrow}_j$ is the standard Pauli matrix.

The ensemble is initialized in the coherent spin state (CSS) along the $x$ axis with the cavity in a vacuum state, i.e. $\langle c^\dagger c\rangle =0$. The initial state is thus
\begin{equation}
\ket{CSS}\otimes \ket{0}=\left({\ket{\downarrow}+\ket{\uparrow}\over \sqrt 2}\right)^N\otimes\ket{0}.
\end{equation}
To elucidate the physical origin of the squeezing, we write the CSS in the basis of Dicke states $\ket{m}$ ($m=-S$, $-S+1$, $\ldots$ $S$) as $\ket{CSS}=\sum_{m=-S}^{S}\sqrt{\binom{2S}{m+S}}\ket{m}$. As shown in Fig.~\ref{Fig1}, the vacuum field only couples $\ket{m}\otimes\ket{0}$ to $\ket{m-1}\otimes\ket{1}$ with a coupling strength $\bra{m}g\sum_{j=1}^N \sigma^+_j\ket{m-1}= g\sqrt{(S+m)(S-m+1)}$. By adiabatically eliminating the cavity operator, we find the AC stark shift due to the vacuum field for the state $\ket{m}$ to be $-\Omega (S+m)(S-m+1) $ with $\Omega=g^2/\Delta$. Therefore, the Hamiltonian governing the evolution of the atomic system is simplified as
\begin{eqnarray}
H'/\hbar&=&\sum_{m=-S}^{S} \left[m\omega_0-\Omega (S+m)(S-m+1)\right]\nonumber\\
& &\times\ket{m}\bra{m} \nonumber\\
&=&-S(S+1)\Omega+\left(\omega_0-\Omega\right)\hat S_z+\Omega\hat S^2_z,
\end{eqnarray}
which is independent of the field operators. The first $S_z$-independent term $-S(S+1)\Omega$ represents a global phase shift. The second term $\left(\omega_0-\Omega\right)\hat S_z$ is a spin precession term due to the transition frequency $\omega_0$ and the cavity-vacuum induced light shift $-\Omega$. The third term $\Omega\hat S^2_z$ is the squeezing term, which commutes with the second term. So in a frame rotating with $\omega_0-\Omega$ we only need to consider the third term, the one-axis twisting Hamiltonian $H_1$ \cite{squeeze}, for the dynamical behavior,
\begin{equation}
H_1/\hbar=\Omega\hat S^2_z.
\end{equation}

In the Heisenberg picture, it is staightforward to obtain the time evolution behavior for $\mean{S_z}$, $\mean{S_y}$, $\mean{S^2_z}$, $\mean{S^2_y}$ and $\mean{S_zS_y+S_yS_z}$ by using the method in Ref. \cite{squeeze},
\begin{eqnarray}
\mean{S_z}&=&0 \\
\mean{S_y}&=&0 \\
\mean{S^2_z}&=&S/2 \\
\mean{S^2_y}&=&S/2+{1\over 2}S(S-{1\over 2})\nonumber \\
& &\times\left[1-\cos^{2S-2}\left(2 \Omega t\right)\right] \\
\mean{S_z S_y+S_y S_z}&=&2S(S-{1\over 2}) \nonumber \\
& &\times\sin\left(\Omega t\right)\cos^{2S-2}\left(\Omega t\right)
\end{eqnarray}

We use the normalized squeezing parameter $\xi(t)=\Delta S^2_{\text{min}}(t)/(S/2)$ to quantify the squeezing \cite{PhysRevA.46.R6797,dynTheory}, where $\Delta S^2_{\text{min}}(t)$ is the minimal variance along an optimum angle. From the above equations it is straightforward to derive
\begin{eqnarray}
\xi(t)&=&\Delta S^2_{\text{min}}(t)/(S/2)\nonumber\\
&=&{1\over S}\bigg[\mean{S^2_z}+\mean{S^2_y}  \nonumber\\
& &-\sqrt{(\mean{S^2_z}-\mean{S^2_y})^2+\mean{S_z S_y+S_y S_z}^2}\bigg] \nonumber \\
&=&1-{1\over 2}(S-{1\over 2})(\sqrt{A^2+B^2}-A),
\end{eqnarray}
where $A=1-\cos^{2S-2}\left(2\Omega t\right)$ and $B=4\sin\left(\Omega t\right)\cos^{2S-2}\left(\Omega t\right)$.  This represents a unitary spin squeezing process, one-axis twisting on the Bloch sphere, which was first introduced by Kitagawa and Ueda \cite{squeeze}. We recognize that vacuum spin squeezing originates in the superradiant coupling of the ensemble to the cavity mode, resulting in an $S_z$-dependent vacuum light shift.

\begin{figure}[htbp]
\begin{center}
\includegraphics[width=3.5in]{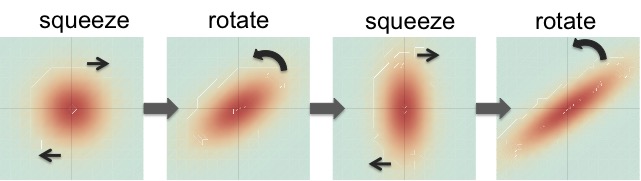}
\caption{Illustration of the exponential squeezing obtained by combining spin squeezing and rotation \cite{GuoGuangCan}. When viewed as a discretized process, each step of squeezing is followed by a rotation. The rotation increases the lever arm $\sqrt{\langle S^2_z \rangle}$ for the next squeezing step, resulting in exponential squeezing with time. When the step size approaches zero, the system is described by the Hamiltonian $H_2/\hbar=\Omega S\hat S_x+\Omega\hat S^2_z$.}
\label{Fig2}
\end{center}
\end{figure}

By adding a rotation about the spin vector, the one-axis twisting can be transformed into an effective two-axis twisting \cite{You1,You2,You3,DuanX}, with improved performance. Fig.~\ref{Fig2} illustrates the mechanism: After the one-axis twisting has produced an uncertainty ellipse whose long axis subtends an angle $\beta$ with the equator, a rotation about the center of the ellipse orients the long axis along $S_z$, thus creating a longer lever arm for the next one-axis twisting. In the limit of a continuous rotation with a rotation speed $\theta=\Omega$ that is matched to the squeezing speed \cite{GuoGuangCan}, the Hamiltonian becomes
\begin{equation}
H_2/\hbar=\Omega S \hat S_x+\Omega \hat S^2_z,
\end{equation}
By applying the bosonic approximation in Ref \cite{nonuniform}, we find the effective Hamiltonian of $H_2/\hbar=\Omega(\hat S^2_z-\hat S^2_y)/2$, whose time evolution can be solved analytically. The squeezing proceeds along the $(\hat z-\hat y)/\sqrt{2}$ direction with an exponential factor $ \exp(-2\Omega t)$, i.e. $\langle S^2_{-\pi/4}\rangle \approx S \exp(-2\Omega t)/2$.

So far, we have ignored two fundamental decoherence processes, namely, photon loss from the cavity at rate $\kappa$ and into free space by atomic emission at rate $\Gamma$ per excited atom. Assuming that the density of atoms is less than $\lambdabar^{-3}$, such that collective (superradiant) emission into free space can be ignored, the free space emission reveals which atom has decayed from $\ket{\uparrow}$ to $\ket{\downarrow}$, and thus destroys the coherence between this atom and the remaining ensemble i.e., it is no longer part of the Dicke ladder depicted in Fig.~\ref{Fig1}. However, this atom is still located inside the optical cavity and will contribute to the final spin measurement of $S_z$ or $S_{\text{min}}$. For a squeezing process of duration $t$, there are on average $\Delta N=N[1-\exp(-\Gamma t)]/2$ atoms transferred from $\ket{\uparrow}$ to $\ket{\downarrow}$ due to spontaneous emission into the free space.  Since each scattering is independent and random, the variance $\delta S^2_{z,\Gamma}$ follows the binomial distribution with $\delta S^2_{z,\Gamma}=\delta{(\Delta N)}^2=S e^{-\Gamma t}(1-e^{-\Gamma t})$. The leakage of a photon from the cavity, on the other hand, does not distinguish between atoms, and maintains the coherence between the atoms while shifting the collective spin $S_z$ down by 1. Shot noise in the leaked photon number then induces a variance in the atomic spin distribution given by $\delta S^2_{z,\kappa}=S\tanh(S\Omega \kappa t/\Delta)[1-\tanh(S\Omega\kappa t/\Delta)]$.

Therefore, the squeezing parameter $\xi(t)$ including both decoherence processes for one-axis twisting can be written as
\begin{eqnarray}
\xi(t)&=&1+{1\over 2}(S-{1\over 2})(A-\sqrt{A^2+B^2})\nonumber \\
& &+2\tanh(S\Omega \kappa t/\Delta)[1-\tanh(S\Omega\kappa t/\Delta)]\nonumber \\
& &+2\exp(-\Gamma t)-2\exp(-2\Gamma t).
\end{eqnarray}

In the limit where decoherence is still small, we find by expansion
\begin{equation}
\xi(t)\approx{\Delta^2\over4 S^2 g^4 t^2}+2{g^2 S\kappa t\over\Delta^2}+2\Gamma t\ge3\left({\Gamma\kappa\over S g^2}\right)^{1/3}=6 \left( N\eta\right)^{-1/3},
\end{equation}
where the single-atom cavity cooperativity is defined as $\eta=4g^2/(\Gamma\kappa)$. If we actively rotate the ensemble continuously with the matched speed, realizing two-axis twisting, then the optimum $\xi$ improves as $(N\eta)^{-1/2}$.

\begin{figure}[htbp]
\begin{center}
\includegraphics[width=2.7in]{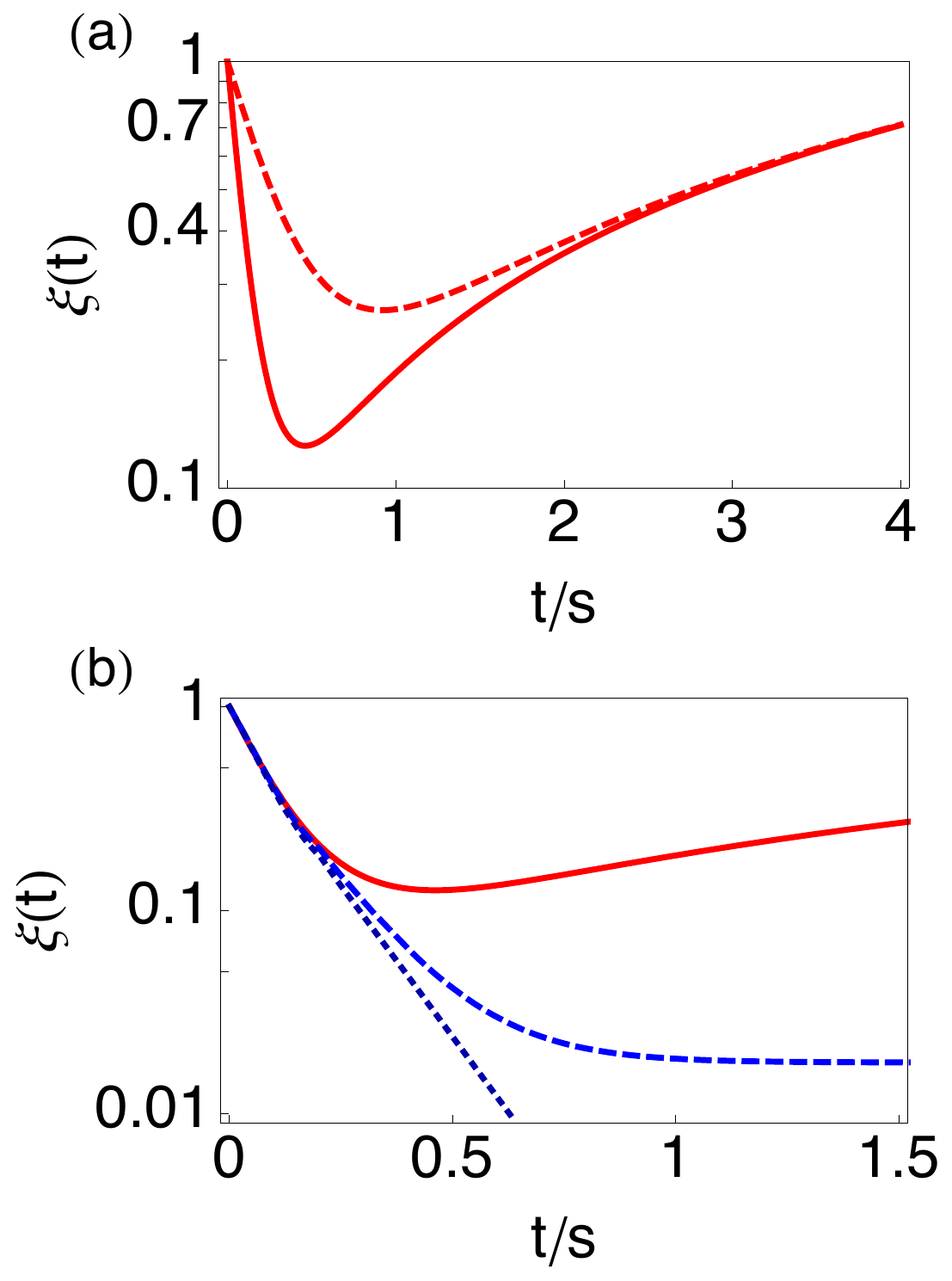}
\caption{(a) Time evolution of the squeezing parameter $\xi(t)$ for $10^4$ $^{171}$Yb atoms ($\Gamma/(2\pi)=7$ mHz) contained in an optical cavity with $\kappa/(2\pi)=100$ kHz and cooperativity $\eta=10$ (red solid line) or $1$ (red dashed line). The detuning is set to $\Delta/(2\pi)=11.2$ MHz for $\eta=10$ (or $\Delta/(2\pi)=3.5$~MHz for $\eta=1$). When $\eta=10$ and $t=0.46$~s, $\xi(t)$ reaches the minimal value, corresponding to 9~dB of squeezing beyond the standard quantum limit. (b) Comparison of one-axis twisting (solid red) and two-axis twisting (dashed blue) when $\eta=10$. The latter is accomplished by appropriate state rotation during the squeezing (see text). Here we include both noise terms, cavity leakage and atomic spontaneous decay. The dotted line shows the evolution of two-axis twisting for a perfect cavity (infinite cooperativity).}
\label{Fig3}
\end{center}
\end{figure}

As a specific example, we consider $10^4$ $^{171}$Yb atoms trapped in an optical cavity with $\Gamma/(2\pi)=7$~mHz and $\kappa/(2\pi)=100$~kHz. In Fig.~\ref{Fig3}(a), we plot the evolution of the squeezing parameter $\xi(t)$ versus time $t$ for both $\eta=10$ and $\eta=1$. In Fig.~\ref{Fig3}(b), we compare the one-axis twisting and effective two-axis twisting when $\eta=10$. Two-axis twisting yields a squeezing of $17.4$~dB after a time $t_s=1$~s, short compared to the excited-state lifetime of 21~s. If the squeezing is extended beyond $t=t_s$, the state remains useful for metrology, but a more complicated procedure to make use of the entanglement-induced increased rotation sensitivity is required \cite{nonG,singledetectionNon}.

We note that the effect of cavity leakage on the atomic state can in principle be suppressed by detecting the photons escaping from the cavity. If the quantum efficiency of the detector is $q$, the noise term can be suppressed by a factor of $1-q$, e.g.
\begin{eqnarray}
\xi(t)&\approx&{\Delta^2\over4 S^2 g^4 t^2}+2(1-q){g^2 S\kappa t\over\Delta^2}+2\Gamma t\nonumber \\
&\ge&3\left[(1-q){\Gamma\kappa\over S g^2}\right]^{1/3}=6 \left[ N\eta/(1-q)\right]^{-1/3}.
\end{eqnarray}
Using the state-of-the-art photon detectors with quantum efficiency above $90\%$ \cite{qd}, an extra factor of 3~dB can be gained for one-axis twisting.

Since one-axis twisting through vacuum spin squeezing does not use any laser light but only an empty cavity relatively far detuned from the atomic transition frequency, it is quite robust. The cavity-enhanced vacuum mode will adiabatically follow any cavity length changes that are slow compared to the cavity linewidth. Since the spin squeezing is a time integral over the instantaneous vacuum light shift, the instantaneous fluctuation of $\Omega(t)$ will cancel in the squeezing phase $\int_0^T dt \Omega(t)$. Considering a squeezing time $t_s$ of 0.073~s, a cavity free spectral range $\nu$ of 5 GHz, a typical standard deviation $\delta\nu$ of 1 MHz, $\Delta/(2\pi)=$11~MHz and the typical noise bandwidth of 10 kHz, the relative error of $\int_0^T dt \Omega(t)$ is $\delta \nu/(\Delta \sqrt{f t_s})=0.004$, and thus is negligible ($10^{-4}$~dB) compared to 9~dB of spin squeezing. To suppress spin squeezing and the vacuum-induced light shift of the clock transition during the Ramsey evolution time of the clock, the cavity can either be mechanically blocked or detuned by $\nu/2$. In the latter case, there are two cavity modes with the same magnitude of detuning but different sign cancelling both the vacuum AC Stark shift and the vacuum squeezing. For a frequency uncertainty $\delta\nu=1$~MHz, the vacuum squeezing is reduced by a factor of ${\Delta\over \nu/2}\times{\delta\nu \over \nu/2}=2\times 10^{-6}$. For the above parameters, $\Omega/(2\pi)=0.16$~mHz, the clock frequency is only shifted by 0.16~mHz during the squeezing time preceding the Ramsey sequence and 0.3~nHz during the Ramsey time (by tuning the detuning), compared to $\omega_0/(2\pi)=518$~THz. Thus, the vacuum spin squeezing does not disturb the accuracy of the optical transition frequency at the $10^{-18}$ level.

In conclusion, we have proposed a new scheme to induce spin squeezing using only an electromagnetic vacuum. By tuning the cavity-enhanced vacuum field relative to a two-level transition with a narrow natural linewidth, the atomic spin distribution spontaneously squeezes beyond the SQL without any external driving field. This offers a simple method to generate spin squeezed states in an optical-transition atomic clocks, where the cavity can also be useful for final state detection at or beyond the SQL \cite{My,sclock,James,20dB}. While the present scheme provides between 10~dB and 20~dB of squeezing under typical condition, we note that for many protocols a moderate amount of squeezing is optimal \cite{PhysRevLett.111.090801,PhysRevA.89.043837,PhysRevLett.92.230801}. Furthermore, by running the clock operation between two vacuum spin squeezing operations of opposite sign (achieved by switching the sign of the cavity detuning), it is possible to realize precision measurement below the SQL without requiring state detection capabilities below the SQL \cite{PhysRevLett.116.053601,Hosten1552}. We also note that for proof-of-principle experiments, one could generate an effective narrow transition using an external laser beam for a Raman transition between two electronic ground states \cite{Thompson74,PhysRevLett.107.063904}.

This work was supported by the NSF, DARPA, NASA and MURI grants through AFOSR and ARO. BB acknowledges support from National Science and Engineering Research Council of Canada.

\bibliographystyle{apsrev4-1}
\bibliography{VISS}

\end{document}